\documentclass[prd, showpacs, preprintnumbers, amsmath, amssymb, nofootinbib, aps, superscriptaddress]{revtex4}

\usepackage{bm,color}
\usepackage{amsfonts}
\usepackage{mathrsfs}
\usepackage{graphicx}
\usepackage{amsmath}
\usepackage{float}
\usepackage{braket}

\begin{document}

\newcommand{\sgn}{\operatorname{sgn}}
\newcommand{\hhat}[1]{\hat {\hat{#1}}}
\newcommand{\pslash}[1]{#1\llap{\sl/}}
\newcommand{\kslash}[1]{\rlap{\sl/}#1}
\newcommand{\lab}[1]{}
% to create labels.
%\newcommand{\iref}[2]{\footnote{\hyperlink{lb:#1}{\textit{$^\spadesuit$#2}}}}
\newcommand{\iref}[2]{}
% reference to other part in this notes.
\newcommand{\sto}[1]{\begin{center} \textit{#1} \end{center}}
\newcommand{\rf}[1]{{\color{blue}[\textit{#1}]}}
% Reference
\newcommand{\eml}[1]{#1}
% Emphasis
\newcommand{\el}[1]{\label{#1}}
% Equation labeling
\newcommand{\er}[1]{Eq.\eqref{#1}}
% Equation Reference
\newcommand{\df}[1]{\textbf{#1}}
% Temporarily replace \textbf
\newcommand{\mdf}[1]{\pmb{#1}}
% Use for vectors etc.
\newcommand{\ft}[1]{\footnote{#1}}
% footnote.
\newcommand{\n}[1]{$#1$}
% Use for numbers etc.
% \newcommand{\cjktext}[1]{\begin{CJK}{GB}{gbsn} #1 \end{CJK}} 
% Language support
\newcommand{\fals}[1]{$^\times$ #1}
% wrong statement
\newcommand{\new}{{\color{red}$^{NEW}$ }}
% update
% \newcommand{\ci}[1]{\cite{#1}}
\newcommand{\ci}[1]{}
\newcommand{\de}[1]{{\color{green}\underline{#1}}}
\newcommand{\ke}{\rangle}
\newcommand{\br}{\langle}
\newcommand{\lb}{\left(}
\newcommand{\rb}{\right)}
\newcommand{\lbk}{\left[}
\newcommand{\rbk}{\right]}
\newcommand{\blb}{\Big(}
\newcommand{\brb}{\Big)}
\newcommand{\nn}{\nonumber \\}
\newcommand{\p}{\partial}
\newcommand{\pd}[1]{\frac {\partial} {\partial #1}}
\newcommand{\cd}{\nabla}
\newcommand{\cc}{$>$}
% ##### ###### ###### ######
\newcommand{\bqa}{\begin{eqnarray}}
\newcommand{\eqa}{\end{eqnarray}}
\newcommand{\bqe}{\begin{equation}}
\newcommand{\eqe}{\end{equation}}
\newcommand{\bay}[1]{\left(\begin{array}{#1}}
\newcommand{\eay}{\end{array}\right)}
\newcommand{\eg}{\textit{e.g.} }
\newcommand{\ie}{\textit{i.e.}, }
\newcommand{\iv}[1]{{#1}^{-1}}
\newcommand{\st}[1]{|#1\ke}
\newcommand{\at}[1]{{\Big|}_{#1}}
\newcommand{\zt}[1]{\texttt{#1}}
\newcommand{\non}{\nonumber}
\newcommand{\m}{\mu}
\newcommand{\ncom}[1]{\textcolor{red}{\bf{[#1]}} }
\newcommand{\nadd}[1]{\textcolor{red}{#1} }
% ##### ###### ###### ######
% Greek Letters
\def\xa{{m}}
\def\xA{{m}}
\def\xb{{\beta}}
\def\xB{{\Beta}}
\def\xd{{\delta}}
\def\xD{{\Delta}}
\def\xe{{\epsilon}}
\def\xE{{\Epsilon}}
\def\xve{{\varepsilon}}
\def\xg{{\gamma}}
\def\xG{{\Gamma}}
\def\xk{{\kappa}}
\def\xK{{\Kappa}}
\def\xl{{\lambda}}
\def\xL{{\Lambda}}
\def\xo{{\omega}}
\def\xO{{\Omega}}
\def\xvp{{\varphi}}
\def\xs{{\sigma}}
\def\xS{{\Sigma}}
\def\xt{{\theta}}
\def\xvt{{\vartheta}}
\def\xT{{\Theta}}
% ##### ###### ###### ######
\def \Tr {{\rm Tr}}
\def\CA{{\cal A}}
\def\CC{{\cal C}}
\def\CD{{\cal D}}
\def\CE{{\cal E}}
\def\CF{{\cal F}}
\def\CH{{\cal H}}
\def\CJ{{\cal J}}
\def\CK{{\cal K}}
\def\CL{{\cal L}}
\def\CM{{\cal M}}
\def\CN{{\cal N}}
\def\CO{{\cal O}}
\def\CP{{\cal P}}
\def\CQ{{\cal Q}}
\def\CR{{\cal R}}
\def\CS{{\cal S}}
\def\CT{{\cal T}}
\def\CV{{\cal V}}
\def\CW{{\cal W}}
\def\CY{{\cal Y}}
\def\BC{\mathbb{C}}
\def\BR{\mathbb{R}}
\def\BZ{\mathbb{Z}}
\def\sA{\mathscr{A}}
\def\sB{\mathscr{B}}
\def\sF{\mathscr{F}}
\def\sG{\mathscr{G}}
\def\sH{\mathscr{H}}
\def\sJ{\mathscr{J}}
\def\sL{\mathscr{L}}
\def\sM{\mathscr{M}}
\def\sN{\mathscr{N}}
\def\sO{\mathscr{O}}
\def\sP{\mathscr{P}}
\def\sR{\mathscr{R}}
\def\sQ{\mathscr{Q}}
\def\sS{\mathscr{S}}
\def\sX{\mathscr{X}}

\def\slz{SL(2,Z)}
\def\slr{$SL(2,R)\times SL(2,R)$ }
\def\ads{${AdS}_5\times {S}^5$ }
\def\adst{${AdS}_3$ }
\def\sun{SU(N)}
\def\ad#1#2{{\frac \delta {\delta\sigma^{#1}} (#2)}}
% for SU(N) SYM
\def\bqf{\bar Q_{\bar f}}
\def\nf{N_f}
\def\sunf{SU(N_f)}

\def\dcirc{{^\circ_\circ}}

\title{Temperatures of  renormalizable quantum field theories in curved spacetime}

\author{Morgan H. Lynch} \email{morgan.lynch@technion.ac.il}
\affiliation{1Department of Electrical Engineering, Technion: Israel Institute of Technology, Haifa 32000, Israel}
\affiliation{Perimeter Institute for Theoretical Physics, 31 Caroline Street North, Waterloo, Ontario N2J 2Y5, Canada}
\affiliation{Leonard E. Parker Center for Gravitation, Cosmology and Astrophysics, University of Wisconsin-Milwaukee,
P.O.Box 413, Milwaukee, Wisconsin USA 53201} 
\author{Niayesh Afshordi} \email{nafshordi@pitp.ca} 
\affiliation{Perimeter Institute for Theoretical Physics, 31 Caroline Street North, Waterloo, Ontario N2J 2Y5, Canada} 
\affiliation{Department of Physics and Astronomy, University of Waterloo, Waterloo, Ontario, N2L 3G1, Canada}

\date{\today}

\begin{abstract}
In this paper we compute the temperature registered by an Unruh-DeWitt detector coupled to a Hadamard renormalizable quantum field in an arbitrary state, moving along an accelerated trajectory in a curved spacetime. For a massless and conformally invariant field, the generalized expression for the temperature is given by the quadratic sum of the 4-acceleration, Raychaudhuri scalar, and renormalized field polarization. We can further find a novel constraint on the renormalized quantum field polarization in relativistic systems that are in global thermal equilibrium. 

\end{abstract}


\maketitle

\section{Introduction}
One of the key insights into quantum gravity provided by quantum field theory in curved spacetime is its prediction that varying geometry can give rise to thermal particle production \cite{parker, hawking, unruh}. This particle production provides a bridge between thermodynamics, quantum theory, and general relativity through the associated temperature. To compute this temperature one can use an Unruh-DeWitt detector \cite{unruh} which requires knowing the Wightman function for the specific field, background geometry, and acceleration in question. Here we use the Hadamard form \cite{marek} of the Wightman function to compute the temperature for a massless conformally coupled Hadamard state. This has the advantage of probing wide classes of quantum fields and geometries which are point-split renormalizable using the Hadamard prescription of subtracting off the singularity structure from the two point function. Using an Unruh-DeWitt detector specifically relies on the Hadamard singularity structure to determine the temperature. To focus on this singularity structure we employ a covariant expansion \cite{folacci} of the Hadamard form evaluated in the quasilocal limit \cite{galley, galley2}. This fully covariant result includes contributions from the background curvature, the quantum state, as well as the proper acceleration of the detector. The resultant expression for the temperature is a generalization of the Page approximation \cite{page} and relates the local temperature to the renormalized vacuum polarization, acceleration, and Raychaudhuri scalar. Furthermore, we can put lower limit on the temperature seen by a local observer, in terms of its trajectory and the local spacetime curvature, independent of the quantum state.

In this paper, Sec. II outlines the conceptual idea behind how the periodicity of a thermal Green's function can be encoded by the singularity structure. With this idea in mind, we examine how the most general statement about the singularity structure can be used to make the most general statement about temperature. Section III employs the covariant and quasilocal expansion of the Hadamard form. The covariant expansion yields polynomials in derivatives of Synge's world function. The quasilocal expansion then writes the resultant expression in terms of powers of a comoving detectors proper time. Included in this expansion are terms that depend explicitly on the local curvature, acceleration, and quantum state.	In Section IV we show how this expansion, when compared to the standard calculation of the Unruh effect, can be used to extract the temperature. The application of our temperature to cosmological settings in Sec. V is used to derive an array of known results as a sanity check. We finalize the analysis in Section VI by examining the regime of validity for the instantaneous temperature to be manifestly real and set a novel constraint on the quantum state.    

\section{encoding temperatures via the singularity structure}

The examination of detector responses in various spacetimes have yielded quite a few examples that are characteristically thermal. The temperatures of these thermal states are typically encoded into the Wightman function of the fields which live and propagate in that spacetime. More specifically, the fact that we have a thermal state is encoded in a periodicity of the Wightman function itself (in imaginary time). As an example, let us consider the general form of the Wightman function for a massless scalar field \cite{bd},

\bqe
G^{\pm}[x,x'] \sim \frac{1}{F(\Delta x , \beta)}.
\eqe

Here the function $F(\Delta x, \beta)$ encodes the spacetime interval $\Delta x^\mu = x^\mu - x'^\mu$ and is periodic under $i\Delta t = i\Delta t+ \beta$ which we identify as the temperature for a thermal state. Since the periodicity is in the denominator, we get singularities when the oscillation crosses zero. As such, with the goal of developing the most general statement of temperature it serves to examine the periodicity encoded in the most general singularity structure of the Wightman function, i.e. the Hadamard form. See Figure 1 for an illustrated example of the idea.  
	
\begin{figure}[H]
\centering  
\includegraphics[scale=.45]{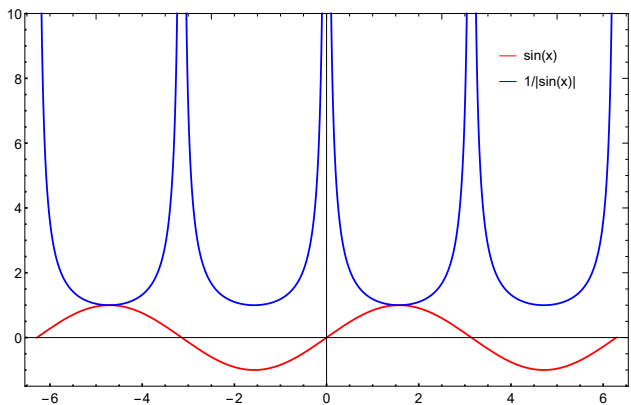}
\caption{Example of the singularity structure of Wightman function in imaginary time.}	
\end{figure}

\section{the covariant and quasilocal expansion of the hadamard form}
Numerous observables in quantum field theory are computed using the two point function and its variants. An observable of particular importance is the expectation value of the quantum energy momentum tensor which is used as the source of the semiclassical Einstein equation. Although the computation of the energy momentum tensor is formally infinite, the Hadamard renormalization prescription, i.e. point splitting, renders the resultant expression finite. This is accomplished by subtracting the singularity structure of the Hadamard form from the two point function used in computing the energy momentum tensor. Another observable of interest is the temperature registered by an Unruh-DeWitt detector. Interestingly enough, it is precisely the singularity structure of the two point function, or more specifically the pole and residue structure, which is needed to compute the temperature. By using the Hadamard form, we are able to use the singularity structure of renormalizable massless and conformally coupled quantum field theories to compute the temperature. The Hadamard form for the symmetrizied Wightman function along a trajectory $x(\tau)$ is given by \cite{galley, galley2, bd}   

\bqa
G^{+}[x(\tau), x(\tau')] \equiv \frac{1}{2} \langle \left\{\phi[x(\tau)],\phi[x(\tau')]\right\} \rangle = \frac{1}{8 \pi^{2}}\lbk \frac{\Delta^{1/2}}{\sigma} +v\ln{(\sigma)} + w   \rbk.
\eqa

Here, the biscalars $v$ and $w$ respectively characterize the quantum field's potential, quantum state, and have the following covariant expansions; $v = \sum_{n} v_{n}(x,x')\sigma^{n}$ and $w = \sum_{n} w_{n}(x,x')\sigma^{n}$. This polynomial expansion is expressed in terms of the covariant derivative of Synge's world function $\nabla_{\alpha}\sigma(x,x') \equiv \sigma_{\alpha}$, where $2\sigma$ is the square of geodesic distance between $x$ and $x'$. Expressing both $\sigma$ and $\sigma_{\alpha}$ utilizing the quasilocal expansion \cite{galley, galley2} we have

\bqa
\sigma &=& -\frac{1}{2}s^{2} - \frac{1}{24}A^{2}s^{4}+{\cal O}(s^5) \non \\
\sigma^{\m} &=& -su^{\m} - \frac{s^{2}}{2}a^{\m}+{\cal O}(s^3).  \label{expansion}
\eqa

Here we have defined the acceleration $A^2 = g_{\m \nu}\frac{D u^{\m}}{d \tau} \frac{D u^{\nu}}{d \tau} = g_{\m \nu}a^{\m}a^{\nu}$ and made use of the identity $2\sigma = \sigma_{\m}\sigma^{\m}$  \cite{Poisson:2011nh}. The parameter $s = \tau' - \tau$ characterizes the proper time of the detector. The overall minus sign indicates that the detector moves along a timelike trajectory normalized via $u^{\alpha}u_{\alpha} = -1$. The covariant expansion of the Van Vleck Moretti determinant is given by \cite{folacci}
\bqa
\Delta^{1/2} &=& 1 + \frac{1}{12}R_{\alpha \beta}\sigma^{\alpha}\sigma^{\beta} + {\cal O} (\sigma^{3/2})
\eqa

Note that if we use the elementary averaging procedure \cite{adler} in the Van Vleck Moretti determinant to exchange the Ricci tensor for the Ricci scalar via $R_{\alpha \beta}\sigma^{\alpha}\sigma^{\beta} \rightarrow R_{\alpha \beta} \frac{1}{2}g^{\alpha \beta}\sigma = \frac{1}{2}R \sigma $, then the expansion is given by $\Delta^{1/2} = 1 + \frac{1}{24}R\sigma$. This allows us to average over the direction of propagation and, more importantly, average over the local curvature in a manner that is independent of the motion, see Fig. 2. This procedure is exact in cases of maximally symmetric spacetimes. 

\begin{figure}[H]
\centering  
\includegraphics[scale=.5]{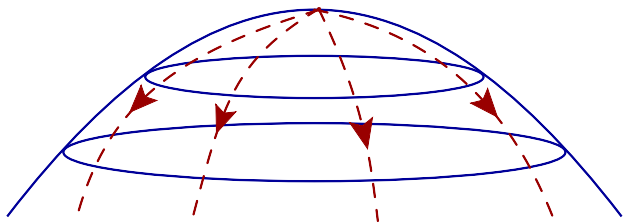}
\caption{Elementary averaging over local curvature as direction of propagation changes.}	
\end{figure}

Finally, if we rewrite the Hadamard form with Synge's \cite{Poisson:2011nh}  world function as a common denominator and expanding to first order in $\sigma$ we have
 
\bqe
G^{+}[x(\tau'), x(\tau)] = \frac{1}{8 \pi^{2}}\lbk \frac{1+ \frac{1}{12}R_{\alpha \beta}\sigma^{\alpha}\sigma^{\beta} +v_{0}\sigma\ln{(\sigma)} + w_{0}\sigma}{\sigma}\rbk .
\eqe

Note that we now have the zeroth order, i.e. coincident limit, terms for the potential and state dependent terms $v_{0}$ and $w_{0}$. The potential is given by $v_{0} = m^{2} + (\xi - \frac{1}{6})R$ while the state dependence can be written in terms of the renormalized vacuum polarization \cite{folacci1} via $w_{0} =  8 \pi^{2} \braket{\phi^2}_{ren} - v_{0}\ln{(\m)}$. Here we introduce a mass scale $\m$ to render the logarithm term in the Hadamard form dimensionless. Finally we have

\bqe
G^{+}[x(\tau'), x(\tau)] = \frac{1}{8 \pi^{2}}\lbk \frac{1+ \frac{1}{12}R_{\alpha \beta}\sigma^{\alpha}\sigma^{\beta} +v_{0}\sigma\ln{(\tilde{\sigma})} + 8 \pi^{2} \sigma \braket{\phi^2}_{ren}}{\sigma}\rbk 
\eqe

Here we have defined the dimensionless world function via $\tilde{\sigma} = \sigma/\m$. For the present analysis we will restrict our fields to be massless and conformally coupled which causes the potential $v_{0}$ to vanish identically and eliminate the logarithm term. As such we drop the term from our current analysis. Finally, utilizing Eqn. (\ref{expansion}) and expanding out to $s^{4}$ in the denominator, we obtain the covariant and quasilocal expansion of the Hadamard form:
\bqe
G^{+}[x(\tau'), x(\tau)]  = \frac{1}{8 \pi^{2}}\lbk \frac{1}{-\frac{1}{2}s^{2} - \frac{1}{24}\lb  A^{2}-R_{\alpha \beta}u^{\alpha}u^{\beta	} +48 \pi^{2}\braket{\phi^2}_{ren}   \rb s^{4} +{\cal O}(s^6)}\rbk. \label{had_wightman}
\eqe

Now that we have developed the Hadamard form to sufficient order, the next step will be to use its singularity structure to compute the temperature registered by an Unruh-DeWitt detector. This is accomplished by looking at the well known example of the Unruh effect, \cite{unruh}. The two point function which is used in this computation is known to give a thermal response and is parametrized by the detector's proper time. By Taylor expanding the two point function about the detectors proper time we are able to directly compare it to our expansion of the Hadamard form and, by inspection, extract the temperature. As such, the covariant expansion ensures we can expand our Hadamard form about the detector's proper time and the quasilocal expansion simultaneously enables us to include contributions from the local curvature as well as the acceleration. The comparison of these two point functions and determination of the their temperatures is carried out in the following section. 

\section{extracting the Generalized Temperature}
The transition rate of an Unruh-DeWitt detector is characterized by the response function \cite{bd}. Formally this is written as the Fourier transform of the two point function, i.e.  

\bqa
\Gamma \propto \int ds ~e^{-i \Delta E s} G^{\pm}[x(\tau'), x(\tau)]. 
\eqa

For a uniformly accelerated observer in Minkowski vacuum, the detector response will be thermal \cite{unruh}. With proper acceleration $a$, the Wightman function which characterizes this uniformly accelerated thermal state is well-known:

\bqe
G_a^{+}[x(\tau'), x(\tau)] = \frac{-a^{2}}{(4\pi)^{2}}\frac{1}{\sinh^2{(as/2)}}.
\eqe

A Wightman function of this form will produce a thermal response. Here, we note the Taylor expansion of this Wightman function will yield the same functional dependence as our Hadamard form. It should be noted that a uniform acceleration in one dimension is a necessary condition for this expansion to yield a thermal response at constant temperature, i.e $\partial_{s} T = 0$ \cite{obadia}. In order to compare with expanded Hadamard form we note the Taylor expansion of the thermal Wightman function	is given by

\bqe
G_a^{+}[x(\tau'), x(\tau)] =  \frac{1}{8 \pi^{2}}\lbk \frac{1}{-\frac{1}{2}s^2 - \frac{1}{24}a^2 s^4+{\cal O}(s^6)}\rbk.
\eqe

Now, we note that the coefficient on the quartic term determines the temperature of the resultant thermal response function. More precisely, the temperature for this thermal state is given by: 

\bqe
T_U^2 = \left(\frac{a}{2 \pi}\right)^2 = -\frac{3}{8\pi^4}\lim_{s \rightarrow 0} \frac{\partial^2}{\partial s^2} \left[\frac{1}{ s^2 G_{a}^{+}[x(\tau'), x(\tau)]}\right], 	\label{unruh}
\eqe

which is known as the Unruh temperature, observed by a constantly accelerated observer in the Minkowski vacuum \cite{unruh}. The limit $s =\tau'-\tau \rightarrow 0$ is significant, as we expect a detector with a large gap $\Delta E \gg a$, which is on for a short period of time $\Delta \tau \ll a^{-1}$, to be only sensitive to this limit. The latter condition can help define an instantaneous notion of temperature, even if acceleration is not constant. With this insight, let's turn to the generic Hadamard Wightman function derived above (Eq. \ref{had_wightman}). We can now {\it define} a generalized notion of {\it instantaneous} temperature:

\bqe
T^2_*(\tau) \equiv  -\frac{3}{8\pi^4}\lim_{s \rightarrow 0} \frac{\partial^2}{\partial s^2} \left[\frac{1} { s^2 G^{+}[x(\tau'), x(\tau)]}\right],
\eqe

for a local observer, moving on an accelerated trajectory, in a generic spacetime, and interacting with a generic Hadamard state of the quantum field. As we saw above, this matches the standard thermodynamic temperature for a thermal state. Plugging Eq. (\ref{had_wightman}) into this definition, we find:

\bqe
T_*^2(\tau) = \frac{1}{(2 \pi)^{2}}\lbk A^{2}- R_{\alpha \beta}u^{\alpha}u^{\beta} +48 \pi^{2}\braket{\phi^2}_{ren} \rbk . \label{t_star}
\eqe

Note we find the generalized temperature is written as the Pythagorean sum of the accelerated temperature $T_{A} = \frac{A}{2 \pi}$, the state temperature $T_{\phi} = 12 \braket{\phi^2}_{ren}$, and a curvature temperature $T_{R} = \frac{\sqrt{- R_{\alpha \beta}u^{\alpha}u^{\beta}}}{2 \pi}$. Moreover, if we had used the elementary averaging procedure the Raychaudhuri scalar would be replaced by the Ricci scalar yielding 

\bqe
\langle T_*^2(\tau)\rangle_{\rm el.~ave.} = \frac{1}{(2 \pi)^{2}}\lbk A^{2}+ \frac{1}{4}R +48 \pi^{2}\braket{\phi^2}_{ren} \rbk .
\eqe

Here, we now have a curvature temperature based off the Ricci scalar $T_{R} = \frac{\sqrt{R}}{4 \pi}$. We should note that the above generalized temperatures are to be considered as an effective temperature that would be registered by a detector and do not necessarily imply a thermal state. Their regime of validity is in the limit of a large detector gap and/or in the near coincidence short time limit. One particular application of this generalized temperature is that it provides insight into the state dependence via the vacuum polarization. Our vacuum polarization is given by 

\bqa
\braket{\phi^2}_{ren}=\frac{1}{12}\lbk T_*^{2} - T_{A}^{2} \rbk + \frac{1}{48\pi^{2}}R_{\alpha \beta}u^{\alpha}u^{\beta}. 
\eqa

In Page's method \cite{page} to compute the renormalized vacuum polarization, he considered a conformally coupled massless scalar in an ultra-static Einstein spacetime where we have $R_{\m \nu} = \Lambda g_{\m \nu}$. For these classes of spacetimes, using either the Ricci or Raychaudhuri scalar, we reproduce celebrated result known as Page's approximation,

\bqa
\braket{\phi^2}_{ren}=\frac{1}{12}\lbk T_*^{2} - T_{A}^{2} \rbk - \frac{\Lambda}{48\pi^{2}}. 
\eqa

\section{Example: Cosmological Backgrounds}

In a cosmological Friedman-Robertson-Walker (FRW) background with scale factor $a(\tau)$, Eq. (\ref{t_star}) takes an interesting form for comoving observers:
\bqe
T^2_{*} (\tau)_{\rm FRW} = 12 \braket{\phi^2}_{ren} + \frac{3}{(2 \pi)^{2}}\frac{\ddot{a}(\tau)}{a(\tau)} ,
\eqe
which combines cosmic accelerations and the state temperature. The 4-acceleration for a comoving observer is zero. Moreover, if we recall that for a massless and conformally coupled scalar field, the vacuum polarization is given by $\braket{\phi^2}_{ren} = -\frac{R}{288 \pi^{2}}$ \cite{castagnino}. The cosmic acceleration can be written in terms of the Hubble constant via $\frac{\ddot{a}(\tau)}{a(\tau)} = \dot{H} + H^{2}$. Recalling that the Ricci scalar in an FRW spacetime is given by $12H^{2}+6\dot{H}$, we find the instantaneous temperature to be,

\bqe
T^2_{*} (\tau)_{\rm FRW} = \frac{1}{(2 \pi)^{2}}\left[ H^{2} +2\dot{H} \right].
\eqe

We can also extend this result to include accelerated observers. The components of the Ricci tensor are given by $R_{00} = 3H^{2} + 3\dot{H}$ and $R_{ii} = -a(t)^{2}(3H^{2} +\dot{H})$ and the normalization of our 4-velocity yields $u_{0}^{2} - a(t)^{2}v^2 = 1$. Then with $\gamma = \frac{d t}{d \tau}$, the boost factor of the observer relative to the cosmological frame, we find the instantaneous temperature measured by an accelerated observer in agreement with \cite{obadia, solveen}. Hence,

\bqe
T^2_{*} (\tau)_{\rm FRW} = \frac{1}{(2 \pi)^{2}}\left[A^{2}+ H^{2} +2\dot{H}\gamma \right].
\eqe

\section{When is temperature real?}

For arbitrary spacetimes, the sign of the curvature correction to the temperature in Eq. (\ref{t_star}) is constrained by the energy conditions. The {\it strong energy condition} requires $R_{\alpha\beta} u^\alpha u^\beta \geq 0$ for time-like $u's$, implying that curvature corrections to $T^2_*$ are always negative:

\bqe
T_*^2(\tau) = T_A^{2}+\frac{1}{(2 \pi)^{2}}\lbk - R_{\alpha \beta}u^{\alpha}u^{\beta} +48 \pi^{2}\braket{\phi^2}_{ren}\rbk \leq T_A^{2} +12\braket{\phi^2}_{ren}.
\eqe   

Furthermore, the weaker {\it null energy condition}  $R_{\alpha\beta} k^\alpha k^\beta \geq 0$, for all null vectors $k^\alpha$ implies that,  even if the curvature correction is positive for an observer (in a spacetime violating the strong energy condition), it will become (arbitrarily) negative for an observer that moves fast enough (unless the null energy condition is saturated). To see this explicitly, let us use Einstein equations for a perfect fluid with density $\rho$ and pressure $p$. Then, Eq. (\ref{t_star}) takes a remarkably simple form:
\bqa
T_*^2(\tau) = T_A^{2}+\frac{1}{(2 \pi)^{2}}\left\{48 \pi^{2}\braket{\phi^2}_{ren}-4\pi G_N \left[ 2(\rho + p) \gamma^2 - \rho +p \right] \right\} \nonumber \\
\leq T_A^{2}+\frac{1}{(2 \pi)^{2}}\left\{48 \pi^{2}\braket{\phi^2}_{ren}-4\pi G_N \gamma^{2}(\rho+3p) \right\} \label{t_perfect}
\eqa
where $G_N$ is the Newton's constant of gravitation, and $\gamma$ is the observer's Lorentz factor in the rest-frame of the fluid. Indeed, for fast observers (large $\gamma$), we get $T^2_*<0$ assuming the null energy condition $\rho+p >0$. %There are two ways to interpret this result: 
%\begin{enumerate} 
%\item One possibility is that a conformally coupled massless field (assumed here) is not expected to directly couple to the Unruh-DeWitt detector, as such an interaction violates conformal symmetry. The detector could instead couple to e.g., field derivatives. We shall defer study of this possibility to future work.  
%\item 
However, one may argue that such fast-moving detectors simply do not register a thermal response function, because they are not interacting with a thermal state. For further discussion, we refer the reader to \cite{schlemmer, bucholtz}.  
%\end{enumerate}

Let us now turn  our focus to thermal gravitationally bound states. We can use Raychaudhuri equation (e.g., \cite{Dadhich:2005qr}) to substitute for $R_{\alpha \beta}u^{\alpha}u^{\beta}$:
\bqe
R_{\alpha \beta}u^{\alpha}u^{\beta} = u^\mu \nabla_\mu (\nabla\cdot u) + 2 (\Sigma^2-\Omega^2) +\frac{1}{3} (\nabla\cdot u)^2- \nabla_\mu a^\mu, \label{ray}
\eqe
where we use the standard definitions for shear and vorticity tensors, $\Sigma$ and $\Omega$, as components of $\nabla_\mu u_\nu$ tensor. A thermal state does not have explicit time dependence, and thus must be in steady state. This implies that:

\bqe
\Sigma^2= \nabla\cdot u =0, {\rm~~ and~~} T_{\rm global} = T_*(x^\mu)/u^0(x^\mu) = {\rm const.},  \label{thermal}
\eqe
where we assume that $u^\mu$'s are the 4-velocities of the observers that see a steady-state thermal state (see e.g., \cite{Rovelli:2010mv}). Now, combining Eq's, (\ref{t_star}) and (\ref{ray}-\ref{thermal}), we find:

\bqe
T^2_{\rm global} = \frac{1}{(2\pi u^0)^2} \left[48 \pi^{2}\braket{\phi^2}_{ren}  + a^\mu a_\mu + \nabla_\mu a^\mu + 2\Omega^2 \right] = {\rm positive ~constant}. \label{vacuum_thermal}
\eqe

This equation provides a novel constraint on the renormalized field polarization in spacetimes with global thermal equilibrium, in terms of the velocity field of its thermal observers:

\bqe
\braket{\phi^2}_{ren}  = \frac{1}{12}(u^0 T_{\rm global})^2 - \frac{1}{48\pi^2}\left(a^\mu a_\mu + \nabla_\mu a^\mu + 2\Omega^2 \right).
\eqe

One final constraint we can make to ensure that we have a well defined temperature is that it is constant with respect to the proper time of the comoving observer, $\partial_{s} T_{\ast} = 0$. This, along with a positive definite, $T_{\ast} > 0$, temperature ensures an exact Taylor series of a thermal Wightman function, e.g. Eqn (10). For spacetimes, trajectories, and quantum states that obey these conditions we will have a well defined thermal response of the detector.

\section{Conclusions}
In this paper, we computed the temperature registered by an Unruh-DeWitt detector accelerating through curved spacetime and coupled to a Hadamard renormalizable massless quantum field. Employing a covariant and quasilocal expansion of the Hadamard form we find a temperature comprised of acceleration, curvature, and quantum state dependent terms. By restricting our geometry to static Einstein spacetimes we can reproduce the Page approximation and an array of cosmological temperatures. Moreover, we found a novel constraint on the renormalized field polarization in spacetimes with global thermal equilibrium.

\section*{Acknowledgments}
This research was supported, in part, by the Leonard E. Parker Center for Gravitation, Cosmology, and Astrophysics, the University of Wisconsin at Milwaukee Department of Physics, and the Perimeter Institute for Theoretical Physics. Research at the Perimeter Institute is supported by the Government of Canada through the Department of Innovation, Science and Economic Development Canada and by the Province of Ontario through the Ministry of Research, Innovation and Science. M.L. was supported in part at the Technion by a Zuckerman Fellowship, the Visiting Graduate Fellowship Program at the Perimeter institute, and a generous scholarship from the UWM Lichtman fund.

\goodbreak

\end{document}